\documentclass[aps,pre,twocolumn,graphicx]{revtex4}

\usepackage[dvips]{graphicx}
\usepackage{color}
\newcommand{\la}{\left\langle}
\newcommand{\ra}{\right\rangle}
\newcommand{\EPL}{Europhys. Lett.~}
\newcommand{\PRL}{Phys. Rev. Lett.~}
\newcommand{\PR}{Phys. Rev.~}
\newcommand{\JCP}{J. Chem. Phys.~}

\newcommand{\PCCP}{Phys. Chem. Chem. Phys.~}

\newcommand{\JPCM}{J. Phys.: Condens. Matter~}
\newcommand{\MP}{Mol. Phys.~}
\newcommand{\JCIS}{J. Coll. Int. Sci.~}
\newcommand{\EPJ}{Eur. Phys. J.~}
\newcommand{\etal}{{\it et~al.}}

\begin{document}

%\preprint{}

% force linebreaks with \\
\title{Electroneutrality and Phase Behavior of Colloidal Suspensions}

\author{A. R. Denton}
\email{alan.denton@ndsu.edu}
\affiliation{Department of Physics, North Dakota State University,
Fargo, ND, 58105-5566}

\date{\today}
\begin{abstract}
Several statistical mechanical theories predict that colloidal suspensions
of highly charged macroions and monovalent microions can exhibit unusual
thermodynamic phase behavior when strongly deionized.  Density-functional,
extended Debye-H\"uckel, and response theories, within mean-field and
linearization approximations, predict a spinodal phase instability of
charged colloids below a critical salt concentration.  Poisson-Boltzmann
cell model studies of suspensions in Donnan equilibrium with a salt reservoir
demonstrate that effective interactions and osmotic pressures predicted by
such theories can be sensitive to the choice of reference system, e.g.,
whether the microion density profiles are expanded about the average
potential of the suspension or about the reservoir potential.  By unifying
Poisson-Boltzmann and response theories within a common perturbative framework,
it is shown here that the choice of reference system is dictated by the
constraint of global electroneutrality.  On this basis, bulk suspensions are
best modeled by density-dependent effective interactions derived from a closed
reference system in which the counterions are confined to the same volume as
the macroions.
Lower-dimensional systems (e.g., monolayers, clusters),
depending on the strength of macroion-counterion correlations, may be governed
instead by density-independent effective interactions tied to an open reference
system with counterions dispersed throughout the reservoir, possibly explaining
observed structural crossover in colloidal monolayers and anomalous
metastability of colloidal crystallites.
\end{abstract}


\maketitle

%\narrowtext
%\twocolumn

%multicols avoids page break:
%\begin{multicols}{2}

%%%%%%%%%%%%%%%%%%%%%%%%%%%%%%%%%%%%%%%%%%%%%%%%%%%%%%%%%%%%%%%%%%%%%%%%%%%
%%%%%%%                       BODY OF TEXT
%%%%%%%%%%%%%%%%%%%%%%%%%%%%%%%%%%%%%%%%%%%%%%%%%%%%%%%%%%%%%%%%%%%%%%%%%%%

\section{Introduction}\label{intro}

%Experiments and simulations:
A variety of experiments have demonstrated the unusual thermodynamic properties
of deionized charge-stabilized colloidal suspensions~\cite{evans99}.
Aqueous suspensions of highly charged macroions and monovalent microions
at sub-millimolar ionic strengths
reportedly can display liquid-vapor coexistence~\cite{tata92},
stable void structures~\cite{ise94,tata97,ise99},
compressed crystals~\cite{ise99,harada99},
metastable crystallites~\cite{grier97}, and
macroion gathering near glass plates~\cite{grier97,muramoto-ito97}.
Such phenomena have been interpreted by some workers~\cite{tata92,ise94,tata97,
ise99} as evidence for pair attraction of like-charged macroions, in seeming
defiance of the classic Derjaguin-Landau-Verwey-Overbeek (DLVO)~\cite{DLVO}
and Poisson-Boltzmann (PB)~\cite{neu99} theories.  Others have attributed
anomalous behavior to nonequilibrium phenomena~\cite{palberg94},
polyelectrolyte impurities~\cite{belloni00}, and many-body
effects~\cite{schmitz99}.  While some  molecular simulations of the primitive
model display macroion aggregation~\cite{allahyarov-damico98,linse-lobaskin99,
lobaskin-linse99,linse99,linse00,rescic-linse01,lobaskin-linse01,
lobaskin-qamhieh03,hynninen05,hynninen07}, such computationally demanding
methods are usually limited to size and charge asymmetries corresponding to
relatively strongly correlated microions and weakly charged macroions.

%Theories:
Several common statistical mechanical theories, including
density-functional~\cite{vRH97,vRDH99,vRE99,zoetekouw_pre06},
extended Debye-H\"uckel~\cite{warren00,chan85,chan01}, and
response~\cite{silbert91,denton99,denton00,denton04,denton06} theories,
predict similarly surprising thermodynamic phase behavior of deionized
suspensions~\cite{int-eq}.
Such coarse-grained theories preaverage the microion degrees of freedom,
reducing the ion mixture to an effective one-component model.  Most 
practical implementations assume a mean-field approximation for the microion
structure, which neglects microion correlations; some form of linearization 
approximation, which ignores nonlinear screening of macroions by microions; 
and a fixed (state-independent) effective macroion charge.
Under these assumptions, the various approaches reduce to variants of 
linearized PB theory, all predicting a screened-Coulomb (Yukawa) effective
pair potential and a one-body volume energy.  Although independent of macroion
coordinates, the volume energy contributes density-dependent terms to the
total free energy that
can drive a spinodal instability of highly charged 
suspensions below a critical salt 
concentration~\cite{vRH97,vRDH99,vRE99,zoetekouw_pre06,warren00,denton06}.
Such unusual phase behavior occurs, however, at parameters as yet inaccessible 
to primitive model simulations and may be qualitatively modified by 
ion correlations and nonlinear screening.

%Linearization and charge renormalization:
Recent studies of colloidal suspensions in
Donnan equilibrium~\cite{donnan11,donnan24} with an electrolyte reservoir
demonstrate that predictions of linearized PB theory can be sensitive to the
choice of reference system~\cite{vongrunberg01,klein01,deserno02,tamashiro03}.
While expansion of the microion densities about the average potential of the
system leads to phase separation at low salt concentrations, expansion about
the reservoir potential strictly predicts phase stability, in qualitative
agreement with nonlinear PB theory~\cite{vongrunberg01,klein01,deserno02,
tamashiro03,trizac03}.  These studies suggest that the predicted phase
separation may be merely a spurious artifact of linearization approximations.

%Why linearize?
Although nonlinear PB theory is often presumed superior to its linearized
form and, by implication, to related linearized theories, it is well to
remember that mean-field theories do not faithfully model nonlinear screening
near highly charged macroions, where counterions are strongly correlated.
Indeed, strong counterion association can renormalize the effective macroion
charge~\cite{alexander84,levin98,tamashiro98,levin01,levin03,levin04,
zoetekouw_prl06}, modifying interparticle interactions and thermodynamic
properties.  Furthermore, nonlinear PB theory is computationally practical
only within cell models~\cite{dobnikar_epl03}, which require computing the
microion distribution around just a single macroion, but thereby completely
neglect macroion correlations.  In contrast, linearized theories predict
analytical expressions for effective interactions that are independent of
any artificial cell geometry, become accurate sufficiently far from the
macroions, and can be input into separate theories or simulations that
incorporate ion correlations and charge renormalization.
Linearized theories thus may play a vital role in multiscale approaches 
to modeling charged colloids.

%Purpose:
The main purpose of this work is to carefully analyze implementations of
linearization approximations in coarse-grained theories of charged colloids.
Connections between PB and response theories are established and exploited to
show that the constraint of global electroneutrality dictates the optimal
choice of reference system.  Theories linearized about a closed reference 
system, whose counterions are confined to the same volume as the macroions, 
are shown to predict density-dependent effective interactions 
and -- assuming a fixed effective macroion charge -- phase instability of 
deionized suspensions.  In contrast, linearization about an 
open reference system, with counterions dispersed throughout the reservoir, 
predicts density-independent effective interactions 
and phase stability, but violates electroneutrality for bulk suspensions.  
Lower-dimensional suspensions (e.g., monolayers, clusters), on the 
other hand, may be modeled best -- for sufficiently weak macroion-counterion 
correlations -- by an open reference system, with important implications for 
thermodynamic properties.

%Outline:
The remainder of the paper is organized as follows.  Section~\ref{model}
first defines the primitive and effective one-component models.
In Sec.~\ref{theory}, effective electrostatic interactions and osmotic
pressures are then derived from response theory and PB theory for closed
and open reference systems.  Corresponding predictions for the equation of
state and phase diagram are presented and contrasted in Sec.~\ref{results}.
Section~\ref{conclusions} closes with a summary and conclusions.

\section{Models}\label{model}

\subsection{Primitive Model}\label{primitive}

\begin{figure}
\includegraphics[width=0.8\columnwidth]{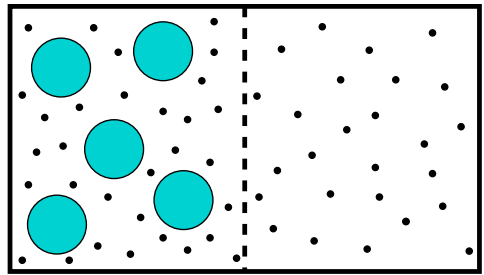}
\caption{\label{donnan} 
Colloidal suspension (left) of macroions (larger circles) and microions
(smaller circles) in Donnan equilibrium with a microion reservoir (right)
via a semipermeable membrane (dashed line), which allows exchange only
of microions.
}
\end{figure}

We consider a charge-stabilized colloidal suspension in Donnan
equilibrium~\cite{donnan11,donnan24} with a microion reservoir
(e.g., electrolyte).  As depicted in Fig.~\ref{donnan}, the suspension and
reservoir are separated by a semipermeable membrane, which allows exchange
of solvent and microions, but not macroions.  The volumes of the suspension
and reservoir are denoted by $V$ and $V_r$, respectively.  Within the
primitive model of charged colloids~\cite{evans99}, the solvent is modeled
as a dielectric continuum, of dielectric constant $\epsilon$, the macroions
as $N_m$ charged hard spheres, of radius $a$ and effective valence $Z$
(charge $-Ze$) and the microions as {\it monovalent} point charges, $N_+$
positive and $N_-$ negative, totaling $N_{\mu}=N_++N_-$ in the suspension.
It is assumed that all ions interact via only bare Coulomb and excluded-volume
pair potentials and have the same dielectric constant as the solvent,
justifying neglect of polarization effects.

The macroions have an average number density $n_m=N_m/V$, their hard cores
occupying a volume fraction $\eta=(4\pi/3)n_m a^3$.  The microions have average
number densities $n_{\pm}=N_{\pm}/V'$ in the free volume $V'=V(1-\eta)$
outside the macroion cores.  The reservoir is assumed to be a 1:1 symmetric
electrolyte with number density of salt ion pairs $n_r$.  Given $Z$ counterions
per macroion and $N_s$ salt ion pairs, the ion numbers are related according
to $N_+=ZN_m+N_s$ and $N_-=N_s$ and the ion densities according to
$n_+=Zn_m/(1-\eta)+n_s$ and $n_-=n_s$, where $n_s=N_s/V'$ is the average
salt density in the free volume.  Overall electroneutrality of the suspension
imposes the constraints $N_+-N_-=ZN_m$ and $n_+-n_-=Zn_m/(1-\eta)$.

\subsection{Effective One-Component Model}\label{ocm}

By averaging over the microion degrees of freedom, the primitive model is
mapped onto a one-component model of pseudomacroions, governed by effective
interactions~\cite{rowlinson84,hansen-lowen00,likos01,levin02,zvelindovsky07}.
This coarse-graining procedure acts on the semigrand partition function
$\langle\langle\exp(-\beta H)\rangle_{\mu}\rangle_m$, where $\la~\ra_m$
denotes a canonical trace over macroion ($m$) coordinates, $\la~\ra_{\mu}$
a grand canonical trace over microion ($\mu$) coordinates, $H$ is the
Hamiltonian, and $\beta\equiv 1/k_BT$ at temperature $T$.  Splitting $H$
into macroion, microion, and macroion-microion interaction terms, according
to $H=H_m+H_{\mu}+H_{m+}+H_{m-}$, and tracing over the microions yields
the semigrand potential,
\begin{equation}
\Omega_{\rm sg}=-k_BT\ln\langle\exp(-\beta H_{\rm eff})\rangle_m,
\label{Omegasg}
\end{equation}
where $H_{\rm eff}=H_m+\Omega_{\mu}$ is an effective one-component Hamiltonian
and
\begin{equation}
\Omega_{\mu}=-k_BT\ln\la\exp\left[-\beta(H_{\mu}+H_{m+}+H_{m-})\right]\ra_{\mu}
\label{Omegamu1}
\end{equation}
is the grand potential of the microions amidst fixed macroions.
Equations (\ref{Omegasg}) and (\ref{Omegamu1}) provide a formal and exact
foundation for both response theory (Sec.~\ref{LRT}), which is based on a
perturbative expansion of $\Omega_{\mu}$ about a reference system, and PB
theory (Sec.~\ref{LPB}), which is based on a density-functional expansion
of the grand potential functional.

\section{Theory}\label{theory}

\subsection{Linear-Response Theory}\label{LRT}

Response theory of charged colloids~\cite{silbert91,denton99,denton00,denton04,
denton06} views the electrostatic potential of the macroions as an ``external"
potential that {\it perturbs} the microions, inducing structure in their
(otherwise uniform) density profiles.  Within this view, the microion grand
potential [Eq.~(\ref{Omegasg})] can be expressed as
\begin{equation}
\Omega_{\mu}=\Omega_0+\int_0^1{\rm d}\lambda\,\left(\la H_{m+}
\ra_{\lambda}+\la H_{m-}\ra_{\lambda}\right),
\label{Omegamu2}
\end{equation}
where $\Omega_0=-k_BT\ln\la\exp(-\beta H_{\mu})\ra_{\mu}$ is the grand
potential of a reference system in which the macroions are uncharged and
the microions are unperturbed, $\lambda$ is a charging parameter, and
$\la~\ra_{\lambda}$ represents a grand canonical ensemble average for
macroion valence $\lambda Z$.
In practice, it proves convenient to add to and subtract from $\Omega_0$
the energy $E_b$ of a uniform, neutralizing background, having charge density
opposite that of the unperturbed microions, and to define
$\Omega_p=\Omega_0+E_b$ as the grand potential of a uniform
(electroneutral) reference microion plasma.

For a sealed suspension whose macroions and microions share the same volume,
the reference system is unambiguous.  For a suspension in Donnan equilibrium,
however, two choices seem to be possible:
\\[0.5ex]
(1) A ``closed" reference system, in which the counterions are confined, with
the macroions, to the suspension.  Independent of reservoir size, the microion
densities in the reference system are then equal to those in the suspension.
\\[0.5ex]
(2) An ``open" reference system, in which the microions are uniformly
distributed throughout the combined volume of the system and the reservoir.
In the limit of an infinite reservoir ($V_r/V\to\infty$), the microion
densities in the reference system equal those in the reservoir.
\\[0.5ex]
\begin{figure}
\includegraphics[width=0.47\columnwidth]{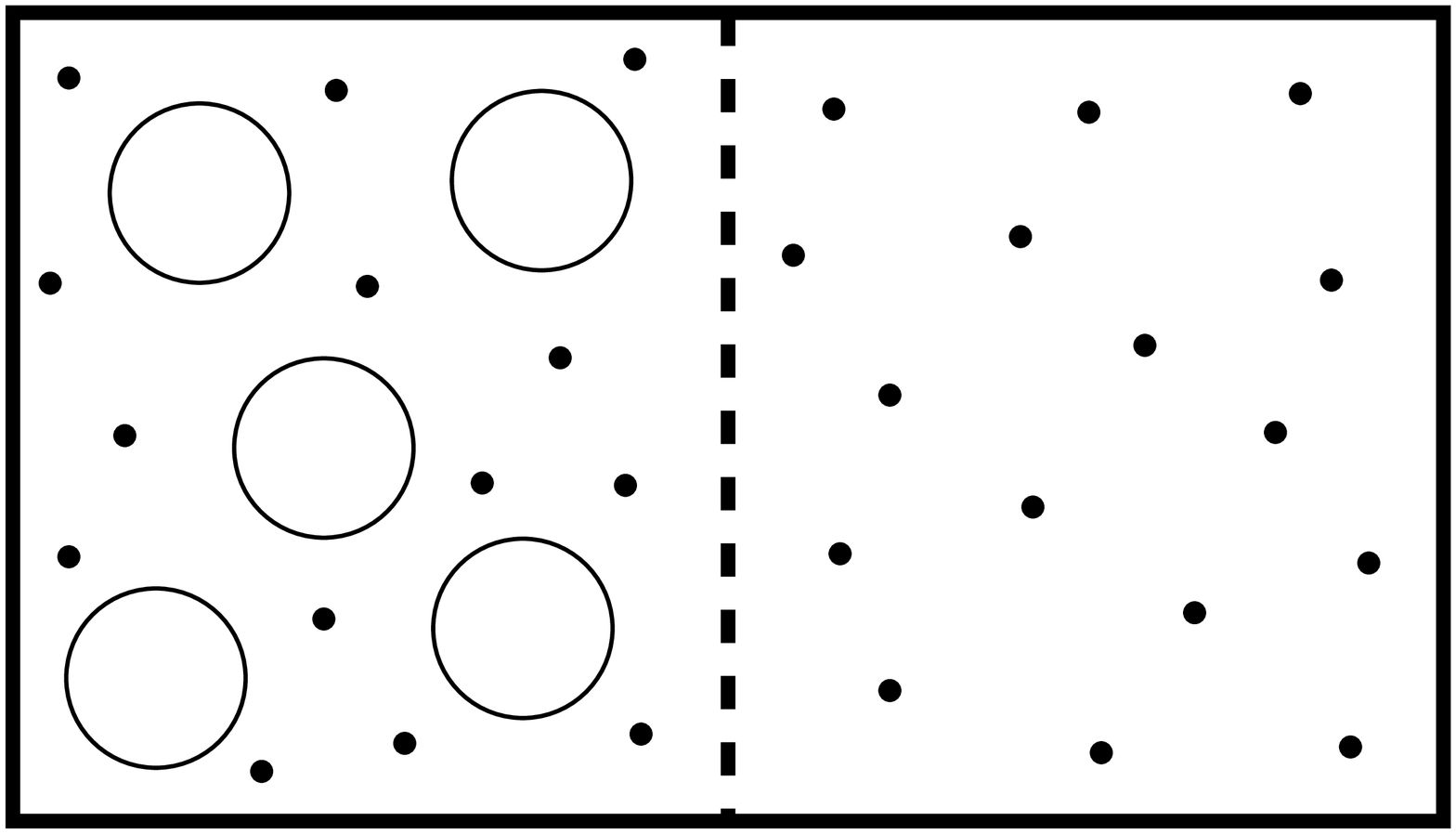}
\hspace*{0.2cm}
\includegraphics[width=0.47\columnwidth]{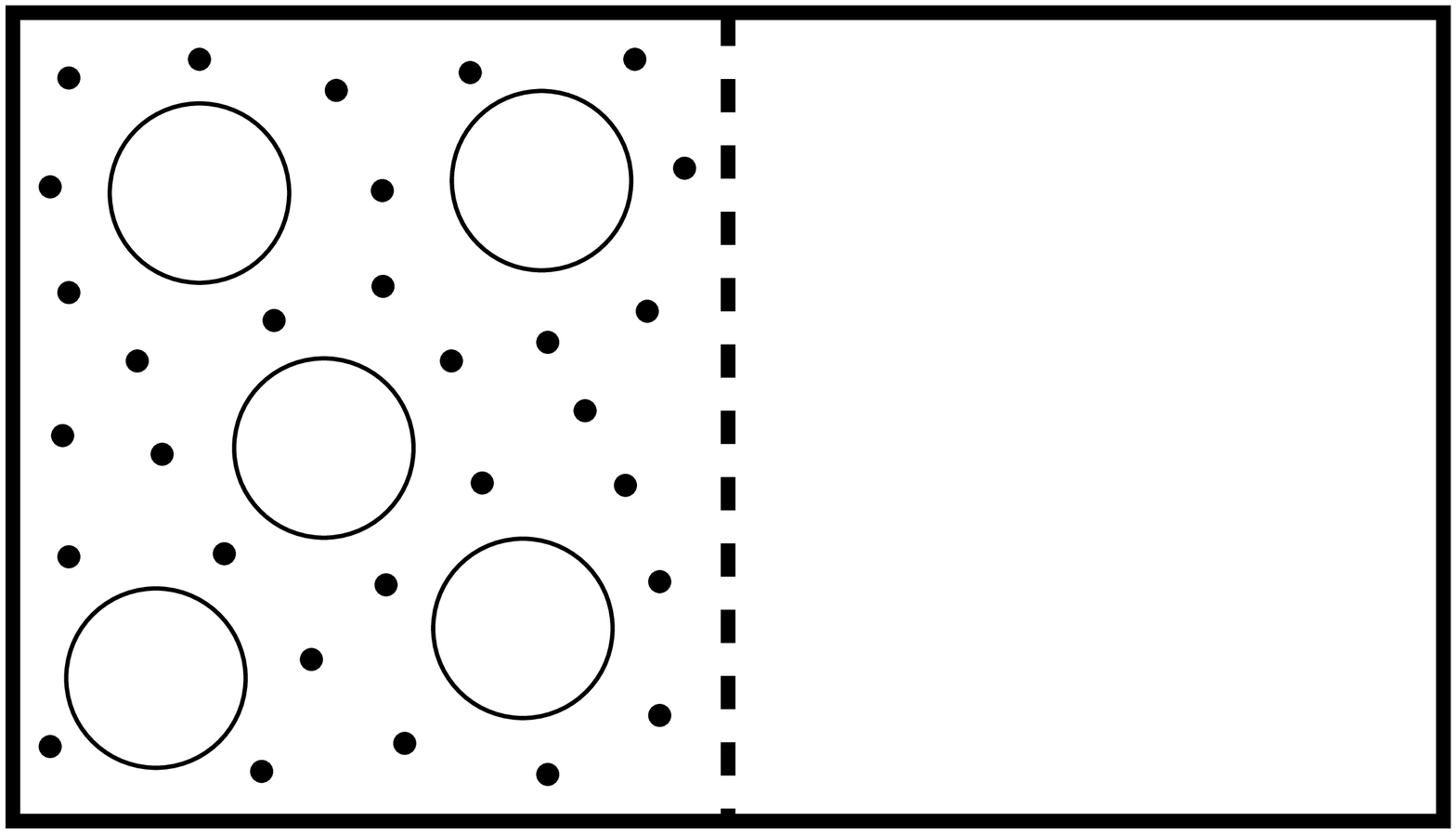}
\caption{\label{reference}
Open (left) and closed (right) reference systems in linear-response theory
of a salt-free colloidal suspension.  Larger circles represent uncharged
(hard-sphere) colloids and smaller circles charged counterions.
}
\end{figure}
The two choices are depicted in Fig.~\ref{reference} in the idealized case
of zero salt concentration.
The second choice may seem most natural, since in the presence of neutral
macroions the counterions would diffuse throughout the available volume
to maximize entropy.  On the other hand, in the real system (with charged
macroions), electroneutrality confines the counterions to the suspension.
Does it matter which reference system is adopted?  By considering each case
in turn, it is shown below that the choice matters, in practice, only for
strongly deionized suspensions, but that the optimal choice then may depend
on system dimensionality.

Linear-response theory~\cite{silbert91,denton99,denton00,denton04,denton06}
is first reviewed, emphasizing the role of the reference system.
The external potential at position ${\bf r}$ is defined as
\begin{equation}
v_{\rm ext}({\bf r})=\int_{V'}{\rm d}{\bf r}'\, v_{m\pm}(|{\bf r}-{\bf r}'|)
n_m({\bf r}'),
\label{vext}
\end{equation}
where $v_{m\pm}(r)=Ze^2/\epsilon r$ ($r>a$) are the macroion-microion
Coulomb pair potentials and $n_m({\bf r})$ is the macroion density.
The macroion-microion interactions then can be expressed as
\begin{eqnarray}
H_{m\pm}&=&\int_{V'}{\rm d}{\bf r}\,v_{\rm ext}({\bf r})n_{\pm}({\bf r})
\nonumber \\[1ex]
&=&\frac{1}{V'}\sum_{\bf k}\hat v_{m\pm}(k)\hat n_m({\bf k})
\hat n_{\pm}(-{\bf k}),
\label{Hmpm}
\end{eqnarray}
where $n_{\pm}({\bf r})$ denote the microion density profiles and the Fourier
transform is defined according to $\hat n_m({\bf k})=\int_{V'}{\rm d}
{\bf r}\exp(-i{\bf k}\cdot{\bf r})n_m({\bf r})$, with inverse
$n_m({\bf r})=(1/V')\sum_{\bf k}\exp(i{\bf k} \cdot{\bf r})\hat n_m({\bf k})$.

Expanding $n_{\pm}({\bf r})$ in a functional Taylor series in powers of
$v_{\rm ext}({\bf r})$ yields~\cite{denton00}
\begin{equation}
\hat n_{\pm}({\bf k})=\chi_{\pm}(k)\hat v_{m\pm}(k)\hat n_m({\bf k})+\cdots,
\quad k\neq 0,
\label{npm}
\end{equation}
where $\chi_{\pm}(k)$ are (partial) linear-response functions, associated with
the microion structure in the reference system, higher-order terms involve
nonlinear response functions and higher powers of the macroion
density~\cite{denton04,denton06}, and the zero-wavevector (long-wavelength)
component is fixed by the normalization condition $\hat n_{\pm}(0)=N_{\pm}$.
Combining Eqs.~(\ref{Omegamu2}), (\ref{Hmpm}), and (\ref{npm}), the
effective Hamiltonian can be written as~\cite{denton99,denton00,denton04}
\begin{equation}
H_{\rm eff}=H_{\rm HS}+E+{1\over{2}}\sum_{i\neq j=1}^{N_m}
v_{\rm eff}(r_{ij})+\cdots,
\label{Heff1}
\end{equation}
where $H_{\rm HS}$ is the hard-sphere (HS) Hamiltonian associated with the
macroion cores, $E$ is a one-body volume energy, $v_{\rm eff}(r)$ is
an effective electrostatic pair potential between macroions whose centers
are separated by $r$, and higher-order terms involve effective many-body
interactions~\cite{denton04,lowen98,goulding99}.
The effective pair potential,
\begin{equation}
v_{\rm eff}(r)=v_{mm}(r)+v_{\rm ind}(r)
\label{veff}
\end{equation}
is the sum of the bare macroion-macroion Coulomb pair potential
$v_{mm}(r)=Z^2e^2/\epsilon r$ and a microion-induced pair potential
$v_{\rm ind}(r)$.  Within a linear-response approximation~\cite{denton00},
\begin{equation}
\hat v_{\rm ind}(k)=\chi(k)[\hat v_{m+}(k)]^2,
\label{vind}
\end{equation}
where $\chi(k)\equiv\chi_+(k)+\chi_-(k)$ is the (total) linear-response
function.  Equations (\ref{Heff1})-(\ref{vind}) are general expressions
for the effective interactions.  More explicit expressions depend on the
specific reference system.

%Closed reference system:
In a {\it closed} reference system, where the unperturbed microions have
average densities $n_{\pm}$, a simple ideal-gas approximation yields
\begin{equation}
\beta\Omega_p=N_+\left[\ln\left(\frac{n_+}{n_r}\right)-1\right]
+N_-\left[\ln\left(\frac{n_-}{n_r}\right)-1\right],
\label{Omegap_closed}
\end{equation}
assuming weakly correlated microions with (electro)chemical
potentials~\cite{note1} fixed by the reservoir:
$\mu_{\pm}=\mu_r=k_BT\ln(n_r\Lambda^3)$, $\Lambda$ being the
microion thermal wavelength.
The one-body volume energy is given by~\cite{denton00}
\begin{eqnarray}
E&=&\Omega_p+\frac{N_m}{2}v_{\rm ind}(0)+N_m(n_+-n_-) \hspace*{1.5cm}
\nonumber \\
&\times&\lim_{k\to 0}
\left[\hat v_{m+}(k)-\frac{1}{2Z}\hat v_{\rm ind}(k)
+\frac{Z}{2}\hat v(k)\right],
\label{E01}
\end{eqnarray}
where $\hat v(k)$ is the Fourier transform of the counterion-counterion
Coulomb pair potential $v(r)=e^2/\epsilon r$.
The linear-response functions are related to the static structure factor
$S(k)$ of the closed reference microion plasma~\cite{HM} via
\begin{equation}
\chi_{\pm}(k)=\mp\beta n_{\pm}S(k)=\mp\frac{\beta n_{\pm}}{1-n_{\mu}\hat c(k)},
\label{chi1}
\end{equation}
where $n_{\mu}=n_++n_-=Zn_m/(1-\eta)+2n_s$ is the total average microion
density in the free volume of the suspension and $\hat c(k)$ is the Fourier
transform of the counterion-counterion direct correlation function.  Within a
random-phase approximation~\cite{HM} for the microion structure, which
neglects all but asymptotically long-ranged microion correlations,
$\hat c(k)=-\beta\hat v(k)$ and the linear-response functions become
\begin{equation}
\chi_{\pm}(k)=\mp\frac{\beta n_{\pm}}{1+\kappa^2/k^2},
\label{chi2}
\end{equation}
where $\kappa=\sqrt{4\pi n_{\mu}\lambda_B}$ is the Debye screening constant
and $\lambda_B=\beta e^2/\epsilon$ is the Bjerrum length.
Taking into account exclusion of the microions from the macroion cores
leads to macroion-microion potentials of the form~\cite{denton00}
\begin{equation}
\hat v_{m\pm}(k)=\mp\frac{4\pi Ze^2}{\epsilon(1+\kappa a)k^2}
\left[\cos(ka)+\kappa\frac{\sin(ka)}{k}\right].
\label{vm+k}
\end{equation}
Substituting Eqs.~(\ref{chi2}) and (\ref{vm+k}) into Eqs.~(\ref{vind}) and
(\ref{E01}) results in explicit analytical expressions for the effective
interactions, namely a screened-Coulomb (Yukawa) effective pair
potential~\cite{denton00}
\begin{equation}
v_{\rm eff}(r)=\frac{Z^2e^2}{\epsilon}\left(\frac{e^{\kappa a}}{1+\kappa a}
\right)^2~\frac{e^{-\kappa r}}{r}, \qquad r\ge 2a
\label{veffr}
\end{equation}
and a volume energy
\begin{equation}
\beta E=\beta\Omega_p-\frac{N_mZ^2}{2}\frac{\kappa\lambda_B}{1+\kappa a}
-\frac{N_mZ}{2}~\frac{n_+-n_-}{n_{\mu}},
\label{E02}
\end{equation}
whose three terms incorporate, respectively, the microion entropy, the
macroion self-energy, and the average microion potential energy.
The latter term is a direct manifestation of the Donnan
effect~\cite{donnan11,donnan24} -- the unequal distribution of microions
between the suspension and reservoir resulting from the impermeability of
the interface to the macroions.  Diffusion of counterions from the suspension
into the reservoir generates an interfacial charge and a resultant electric
field that pulls the counterions back to maintain global electroneutrality
of the suspension.  The electrostatic potential of the suspension is thereby
shifted, by the Donnan potential $\Psi_D$, relative to that of the reservoir
to equalize the microion chemical potentials.  In terms of $\Psi_D$, the final
term in Eq.~(\ref{E02}) can be expressed as $(n_+-n_-)e\Psi_D/2$, which is
simply the work required to move microions from the reservoir
(at zero potential) to the suspension (at potential $\Psi_D$).

It is vital to note that the effective pair potential [Eq.~(\ref{veffr})]
and volume energy [Eq.~(\ref{E02})] derived from the closed reference system
depend implicitly (via $\kappa$) on the average microion densities in
the suspension.  The electroneutrality constraint then imposes a dependence
of $v_{\rm eff}(r)$ and $E$ on the macroion density.  As discussed below
in Sec.~\ref{results}, this density dependence entails an effective
many-body cohesion that can profoundly impact the bulk phase behavior of
counterion-dominated suspensions.

%Open reference system:
In an {\it open} reference system, the unperturbed microions are not confined
with the macroions, but instead uniformly fill the combined free volume of the
suspension and reservoir.  The effective interactions are now still described
by Eqs.~(\ref{veffr}) and (\ref{E02}), but with the average microion densities
in the suspension $n_{\pm}$ replaced by the average microion densities in the
suspension + reservoir, i.e., $(n_{\pm}+n_rV_r/V')/(1+V_r/V')$.  In the limit
of an infinite reservoir ($V_r/V'\to\infty$), the reference microion densities
approach $n_r$ and the reference microion plasma grand potential
[Eq.~(\ref{Omegap_closed})] $\beta\Omega_p\to -2n_rV'$.
In this same limit, the linear-response functions, now associated with the
microion structure of the reservoir, become [cf. Eq.~(\ref{chi2})]
\begin{equation}
\chi_{\pm}^{(r)}(k)=\mp\frac{\beta n_r}{1+\kappa_r^2/k^2},
\label{chi3}
\end{equation}
where $\kappa_r=\sqrt{8\pi n_r\lambda_B}$ is the {\it reservoir} screening
constant.  Likewise, the effective pair potential retains the Yukawa form
[Eq.~(\ref{veffr})] with $\kappa$ replaced by $\kappa_r$ and the
volume energy reduces to [cf. Eq.~(\ref{E02})]
\begin{equation}
\beta E_r=-2n_rV'-\frac{N_mZ^2}{2}\frac{\kappa_r\lambda_B}{1+\kappa_r a}.
\label{Evol-open}
\end{equation}
Note that the electroneutrality term [final term of Eq.~(\ref{E02})] vanishes,
since in the infinite-reservoir limit the reference microion plasma contains
equal densities of positive and negative microions, the counterions having
``evaporated" into the reservoir.

Thus, when applied to a suspension in Donnan equilibrium with a microion
reservoir, linear-response theory predicts nontrivial dependence of the
effective interactions on macroion density only when the effective Hamiltonian
is perturbed about a closed reference system.  If the perturbative expansion
is performed about an open reference system, the volume energy --
in the case of an infinite reservoir -- is independent of macroion density
and hence does not influence thermodynamics.  Clearly the distinction between
reference systems can have practical relevance only if the counterion
concentration is not overwhelmed by the salt ion concentration.
Section~\ref{results} identifies a physical criterion for selecting
the optimal reference system and explores implications for phase stability
of deionized suspensions.

\subsection{Linearized Poisson-Boltzmann Theory}\label{LPB}

An alternative, and widely studied, approach to modeling charged colloidal
suspensions and polyelectrolyte solutions is the Poisson-Boltzmann theory,
here briefly reviewed with emphasis on connections to response theory and
the role of the reference system.  Poisson-Boltzmann theory can be derived
from the microion grand potential~\cite{lowen93}
\begin{equation}
\Omega_{\mu}[n_+({\bf r}),n_-({\bf r})]={\cal F}_{\mu}-\mu_+N_+-\mu_-N_-,
\label{Omegamu3}
\end{equation}
regarded as a functional of the nonuniform microion number densities
$n_{\pm}({\bf r})$ in the external potential $v_{\rm ext}({\bf r})$ of the
macroions [Eq.~(\ref{vext})].  The microion Helmholtz free energy functional
separates naturally, according to
\begin{equation}
{\cal F}_{\mu}[n_+({\bf r}),n_-({\bf r})]={\cal F}_{\rm id}
+{\cal F}_{\rm ext}+{\cal F}_{\rm ex},
\label{Fmu1}
\end{equation}
into an ideal-gas free energy
\begin{eqnarray}
{\cal F}_{\rm id}&=&k_BT\int_{V'}{\rm d}{\bf r}\,\left(n_+({\bf r})
\{\ln[n_+({\bf r})\Lambda^3]-1\}\right. \nonumber \\[1ex]
&+&\left.n_-({\bf r})\{\ln[n_-({\bf r})\Lambda^3]-1\}\right),
\label{Fid1}
\end{eqnarray}
due to the microion entropy, an external free energy
\begin{equation}
{\cal F}_{\rm ext}=\int_{V'}{\rm d}{\bf r}\,v_{\rm ext}({\bf r})
[n_+({\bf r})-n_-({\bf r})],
\label{Fext}
\end{equation}
associated with microion-macroion interactions, and an excess free energy
${\cal F}_{\rm ex}$, due to microion-microion interactions.
In a mean-field (Hartree) approximation, which neglects all interparticle
correlations, the excess free energy can be written as
\begin{eqnarray}
{\cal F}_{\rm ex}&=&\frac{1}{2}\int_{V'}{\rm d}{\bf r}\,
\int_{V'}{\rm d}{\bf r}'\,v(|{\bf r}-{\bf r'}|)
[n_+({\bf r})-n_-({\bf r})] \nonumber \\[1ex]
&\times&[n_+({\bf r}')-n_-({\bf r}')].
\label{Fex1}
\end{eqnarray}

Combining Eqs.~(\ref{Omegamu3})-(\ref{Fex1}) and minimizing
$\Omega_{\mu}[n_+({\bf r}),n_-({\bf r})]$ with respect to $n_{\pm}({\bf r})$,
for a given macroion density, leads to Boltzmann distributions for
the equilibrium microion densities:
\begin{equation}
n_{\pm}({\bf r})=n_r\exp[\mp\beta e\Psi({\bf r})],
\label{npm1}
\end{equation}
where the electrostatic potential $\Psi({\bf r})$ is defined via
\begin{equation}
e\Psi({\bf r})=\int_{V'}{\rm d}{\bf r}'\,v(|{\bf r}-{\bf r'}|)n({\bf r}'),
\label{Psi}
\end{equation}
with $n({\bf r})\equiv n_+({\bf r})-n_-({\bf r})-n_f({\bf r})$ the total
number density of all charges, including those fixed on the macroion surfaces
$n_f({\bf r})$.
Equation (\ref{npm1}) and the exact Poisson equation
\begin{equation}
\nabla^2\Psi({\bf r})=-\frac{4\pi e}{\epsilon}n({\bf r})
\label{Poisson1}
\end{equation}
together determine $\Psi({\bf r})$ and $n_{\pm}({\bf r})$
under prescribed boundary conditions.
The total free energy functional (for fixed macroions),
${\cal F}=E_{mm}+{\cal F}_{\mu}$, which includes the macroion-macroion
Coulomb interaction energy, $E_{mm}=\sum_{i<j}v_{mm}(r_{ij})$, is given by
\begin{equation}
{\cal F}={\cal F}_{\rm id}+\frac{e}{2}\int_{V'}{\rm d}{\bf r}\,
\Psi({\bf r})n({\bf r})
\label{FPB1}
\end{equation}
or, within a cell model approximation,
\begin{equation}
{\cal F}={\cal F}_{\rm id}+N_m\frac{\epsilon}{8\pi}\int_{\rm cell}
{\rm d}{\bf r}\,|\nabla\Psi({\bf r})|^2,
\label{FPB2}
\end{equation}
using Eq.~(\ref{Poisson1}) and assuming vanishing electric field
($\nabla\Psi=0$) at the cell boundary.

More generally, independent of any cell geometry, and for any fixed macroion
distribution, the microion free energy functional can be expressed as
\[
{\cal F}_{\mu}={\cal F}_{\rm id}
+\frac{1}{V'}\sum_{\bf k}\hat v_{m+}(k)\hat n_m({\bf k})
[\hat n_+(-{\bf k})-\hat n_-(-{\bf k})]
\]
\vspace{-0.5cm}
\begin{equation}
+\frac{1}{2V'}\sum_{\bf k}\hat v(k)[\hat n_+({\bf k})-\hat n_-({\bf k})]
[\hat n_+(-{\bf k})-\hat n_-(-{\bf k})].
\label{FPB3}
\end{equation}
A systematic scheme for further approximations is based on expansion of the
microion density profiles [Eq.~(\ref{npm1})] about a reference potential
$\Psi_0$ and a functional Taylor-series expansion of the
ideal-gas free energy [Eq.~(\ref{Fid1})] about reference densities
$n_{\pm}^{(0)}$.  In linearized PB theory, the expansions of
$n_{\pm}({\bf r})$ are truncated at linear order in the potential deviation,
\begin{equation}
n_{\pm}({\bf r})=n_{\pm}^{(0)}\{1\mp\beta e[\Psi({\bf r})-\Psi_0]\},
\label{npm2}
\end{equation}
and the expansion of ${\cal F}_{\rm id}$ is truncated at quadratic order
in the density deviations,
\begin{eqnarray}
{\cal F}_{\rm id}&=&
k_BT\sum_{i=\pm}\left\{N_i\ln\left(n_i^{(0)}\Lambda^3\right)-V'n_i^{(0)}
\phantom{\frac{1}{2n_i^{(0)}}}\right.
\nonumber \\
&+&\left.\frac{1}{2n_i^{(0)}}\int_{V'}{\rm d}{\bf r}\,
\left[n_i({\bf r})-n_i^{(0)}\right]^2\right\},
\label{Fid2}
\end{eqnarray}
where $n_{\pm}^{(0)}\equiv n_r\exp(\mp\beta e\Psi_0)$ are the reference
microion densities.
Analogous to the choice in linear-response theory between closed and open
reference systems, here two choices of reference potential and corresponding
microion densities are apparent:
\\[0.5ex]
(1) The average potential and microion densities in the suspension,
$\Psi_0=\overline\Psi$ and $n_{\pm}^{(0)}=n_{\pm}$.
\\[0.5ex]
(2) The average potential and microion densities in the reservoir,
$\Psi_0=0$ and $n_{\pm}^{(0)}=n_r$.
\\[0.5ex]
Although the second choice is most commonly assumed, the first has been
advocated~\cite{vongrunberg01,klein01,deserno02} as better exploiting the
supposedly weak deviations,
[$\Psi({\bf r})-\overline\Psi$] and [$n_{\pm}({\bf r})-n_{\pm}$].

Expanding $n_{\pm}({\bf r})$ first about $\Psi_0=\overline\Psi$, and combining
the Fourier transform of the linearized expansion with that of Eq.~(\ref{Psi}),
\begin{equation}
e\hat\Psi({\bf k})=\hat v_{m+}(k)\hat n_m({\bf k})+
\hat v(k)[\hat n_+({\bf k})-\hat n_-({\bf k})],
\label{Poisson2}
\end{equation}
the microion densities take on the form
\begin{equation}
\hat n_{\pm}({\bf k})=\mp n_{\pm}\beta e\hat\Psi({\bf k})
=\chi_{\pm}(k)\hat v_{m\pm}(k)\hat n_m({\bf k}),~~k\neq 0,
\label{npm3}
\end{equation}
where $\chi_{\pm}(k)$ are identical to the linear-response functions
that appear in Eqs.~(\ref{npm}) and (\ref{chi2}).
Equation~(\ref{npm3}) implies that
\begin{equation}
\hat n_+({\bf k})-\hat n_-({\bf k})=\chi(k)\hat v_{m+}(k)
\hat n_m({\bf k}), \qquad k\neq 0.
\label{npm4}
\end{equation}
From the inverse transform of Eq.~(\ref{npm3}), using Eqs.~(\ref{chi2})
and (\ref{vm+k}), the electrostatic potential around an isolated macroion
is then given by
\begin{equation}
\Psi(r)=-\frac{Ze}{\epsilon}\frac{e^{\kappa a}}{1+\kappa a}
~\frac{e^{-\kappa r}}{r}, \qquad r\ge a,
\label{Psir}
\end{equation}
which is identical to the solution of the Poisson equation
[Eq.~(\ref{Poisson1})] with boundary conditions
$\Psi'(r)|_{r=a}=-Ze/\epsilon a^2$ and $\Psi(r)\to 0$ as $r\to\infty$.
The corresponding expansion of ${\cal F}_{\rm id}$ about the average microion
densities becomes
\begin{eqnarray}
{\cal F}_{\rm id}&=&k_BT\sum_{i=\pm}\left\{N_i[\ln(n_i\Lambda^3)-1]
\phantom{\sum_{{\bf k}\neq 0}}\right.
\nonumber \\
&+&\left.\frac{1}{2n_iV'}\sum_{{\bf k}\neq 0}\hat n_i({\bf k})
\hat n_i(-{\bf k})\right\},
\label{Fid3}
\end{eqnarray}
where the $k=0$ term vanishes via normalization:
$\int_{V'}{\rm d}{\bf r}\, n_{\pm}({\bf r})=n_{\pm}V'$.
Combining Eqs.~(\ref{FPB3}), (\ref{npm3}), and (\ref{Fid3}) then yields
the microion Helmholtz free energy (equilibrium value of ${\cal F}_{\mu}$)
for fixed macroions,
\newpage
\begin{eqnarray}
F_{\mu}&=&k_BT\sum_{i=\pm}N_i[\ln(n_i\Lambda^3)-1]
\nonumber \\
&+&\frac{1}{2V'}\sum_{{\bf k}\neq 0}
\hat v_{\rm ind}(k)\hat n_m({\bf k})\hat n_m(-{\bf k})+(n_+-n_-)
\nonumber \\
&\times&\lim_{k\to 0}\left[\frac{1}{2}(N_+-N_-)\hat v(k)+N_m\hat v_{m+}(k)
\right],
\label{FPB4}
\end{eqnarray}
the $k=0$ terms being determined by the electroneutrality constraint.
Significantly, this free energy generates {\it precisely} the same
effective Hamiltonian, $H_{\rm eff}=H_{\rm HS}+F-N_{\mu}\mu_r$, as derived
above (Sec.~\ref{LRT}) from response theory perturbed about a closed
reference system [Eqs.~(\ref{Heff1})-(\ref{E01})].

Expanding $n_{\pm}({\bf r})$ now about the reservoir potential ($\Psi_0=0$),
the Fourier transform of the linearized microion densities combined with
Eq.~(\ref{Poisson2}) yields
\begin{equation}
\hat n_{\pm}({\bf k})=\mp n_r\beta e\hat\Psi({\bf k})
=\chi_{\pm}^{(r)}(k)\hat v_{m\pm}^{(r)}(k)\hat n_m({\bf k}),~~k\neq 0,
\label{npm5}
\end{equation}
from which
\begin{equation}
\hat n_+({\bf k})-\hat n_-({\bf k})=\chi^{(r)}(k)\hat v_{m+}^{(r)}(k)
\hat n_m({\bf k}), \qquad k\neq 0,
\label{npm6}
\end{equation}
where $\chi_{\pm}^{(r)}(k)$ are the linear-response functions of
Eq.~(\ref{chi3}), $\chi^{(r)}(k)\equiv\chi_+^{(r)}(k)-\chi_-^{(r)}(k)$,
and $\hat v_{m\pm}^{(r)}(k)$ are given by Eq.~(\ref{vm+k}), with $\kappa$
replaced by $\kappa_r$.
The expansion of ${\cal F}_{\rm id}$ about the reservoir microion density
becomes
\begin{eqnarray}
\beta{\cal F}_{\rm id}&=&N_{\mu}[\ln(n_r\Lambda^3)-1]-n_rV'
\nonumber \\[1ex]
&+&\frac{1}{2n_rV'}\sum_{i=\pm}\sum_{\bf k}\hat n_i({\bf k})\hat n_i(-{\bf k}),
\label{Fid5}
\end{eqnarray}
the $k=0$ term now nonvanishing.
Substituting Eqs.~(\ref{npm5})-(\ref{Fid5}) into Eq.~(\ref{FPB3}) then yields
\begin{eqnarray}
\beta F_{\mu}&=&N_{\mu}[\ln(n_r\Lambda^3)-1]-n_rV'
+\frac{(N_++N_-)^2}{4n_rV'}
\nonumber \\
&+&\frac{\beta}{2V'}\sum_{\bf k}\hat v_{\rm ind}^{(r)}(k)
\hat n_m({\bf k})\hat n_m(-{\bf k}),
\label{FPB5}
\end{eqnarray}
where $\hat v_{\rm ind}^{(r)}(k)\equiv\chi^{(r)}(k)[\hat v_{m+}^{(r)}(k)]^2$
is an induced potential that depends (via $\kappa_r$) on only the reservoir
microion density $n_r$.
This free energy corresponds to essentially the same effective Hamiltonian
as derived from response theory perturbed about the open (reservoir)
reference system, including the same effective pair potential
[Eq.~(\ref{veffr}) with reservoir screening constant $\kappa_r$], but
a slightly different volume energy [cf. Eq.~(\ref{Evol-open})],
\begin{equation}
\beta E_r=V'\left(\frac{n_{\mu}^2}{4n_r}-n_{\mu}-n_r\right)
-\frac{N_mZ^2}{2}\frac{\kappa_r\lambda_B}{1+\kappa_r a}.
\label{Evol-open-PB}
\end{equation}

Two main conclusions follow from the above derivations.  First, linearization
about the average potential and microion densities of the suspension leads to
effective interactions that depend on macroion density, while linearization
about the reservoir yields density-independent interactions.  Second,
linear-response and linearized PB theories are formally equivalent~\cite{note2},
with direct correspondences between closed/open and suspension/reservoir
reference systems.
In Sec.~\ref{results}, these connections are exploited to bolster previous
arguments~\cite{vongrunberg01,klein01,deserno02,tamashiro03} for linearizing
PB theory about the average potential of the suspension, rather than that
of the reservoir.

\subsection{Thermodynamic Properties}\label{thermo}

%Thermodynamics:
Despite the presence of three ion species in the real suspension,
thermodynamic properties of the effective one-component model depend on
the chemical potentials of only pseudomacroions $\mu_m$ and salt ion pairs
$\mu_s$, since electroneutrality permits exchange of ions only in
electroneutral units.
A suspension in Donnan equilibrium with a salt reservoir, at fixed salt
chemical potential $\mu_s=2\mu_r=2k_BT\ln(n_r\Lambda^3)$, has a
total pressure $p=n_m\mu_m-\omega$ and a pseudomacroion chemical potential
$\mu_m=(\partial\omega/\partial n_m)_{\mu_s}$, where $\omega\equiv\Omega/V$.
The semigrand potential density of the suspension can be expressed as
\begin{equation}
\omega(n_m,\mu_s)=f_{\rm eff}+\varepsilon,
\label{omega}
\end{equation}
where $f_{\rm eff}$ is the free energy density of the one-component system
interacting via the Yukawa effective pair potential [Eq.~(\ref{veffr})],
with screening constant $\kappa$ or $\kappa_r$, and $\varepsilon=E/V$
[Eq.~(\ref{E02})] or $E_r/V$ [Eq.~(\ref{Evol-open})] for closed or open
reference systems, respectively.  The salt density in the suspension is
determined by solving for $n_s$ the implicit relation
$\mu_s=(\partial f/\partial n_s)_{n_m}$, where
\begin{equation}
f(n_m,n_s)=\omega+\mu_s n_s
\label{f}
\end{equation}
is the total free energy density of the suspension~\cite{note3}.

The dependence of effective interactions on the choice of reference system
has important implications for the osmotic pressure of deionized suspensions.
Substituting the appropriate volume energies into Eq.~(\ref{omega}) yields,
for the closed reference system,
\begin{equation}
\beta p=n_{\mu}+n_m+\beta p_{\rm ex}(n_{\mu})
-\frac{Z(n_+-n_-)\kappa\lambda_B}{4(1+\kappa a)^2},
\label{p-closed}
\end{equation}
and for the open reference system and an infinite reservoir,
\begin{equation}
\beta p=2n_r+n_m+\beta p_{\rm ex}(n_r),
\label{p-open}
\end{equation}
where
\begin{equation}
p_{\rm ex}=n_m\left(\frac{\partial f_{\rm ex}}{\partial n_m}\right)_{N_s/N_m}
-f_{\rm ex}
\label{pex}
\end{equation}
is the excess pressure due to effective macroion-macroion pair interactions
(with respective screening constant $\kappa$ or $\kappa_r$) and
$f_{\rm ex}=F_{\rm ex}/V$ is the excess free energy density.  The first two
terms on the right sides of Eqs.~(\ref{p-closed}) and (\ref{p-open}) are the
pressure contributions from microion and macroion translational entropy.
The final term in Eq.~(\ref{p-closed}), notably absent from Eq.~(\ref{p-open}),
results from the density dependence of the microion-macroion interactions
and thus is a manifestation of electroneutrality.  For the open reference system
and a finite reservoir, Eq.~(\ref{p-closed}) applies with the simple reassignment
$n_{\pm}\to (n_{\pm}+n_rV_r/V')/(1+V_r/V')$, i.e.,
average microion densities in suspension + reservoir.

%Variational Free Energy Theory:
The excess free energy density can be accurately approximated by a
variational method~\cite{vRH97,vRDH99,vRE99,denton06}
based on first-order thermodynamic perturbation theory with a
hard-sphere reference system~\cite{HM}:
\begin{eqnarray}
&&f_{\rm ex}(n_m,n_s)=\min_{(d)}\left\{f_{\rm HS}(n_m,n_s;d)+2\pi n_m^2
\phantom{\int_d^{\infty}}\right.
\nonumber \\[1ex]
&\times&\left.
\int_d^{\infty}{\rm d}r\, r^2g_{\rm HS}(r,n_m;d)v_{\rm eff}(r,n_m,n_s)
\right\},
\label{fex}
\end{eqnarray}
where the effective hard-sphere diameter $d$ is the variational parameter
and $f_{\rm HS}(n_m,n_s;d)$ and $g_{\rm HS}(r,n_m;d)$ are the excess free
energy density and (radial) pair distribution function, respectively, of
the HS fluid, computed here from the essentially exact Carnahan-Starling and
Verlet-Weis expressions~\cite{HM}.  From the Gibbs-Bogoliubov
inequality~\cite{HM}, minimization with respect to $d$ generates a
least upper bound to the free energy.

%Phase diagram:
The phase diagram can be computed from a (Maxwell) common-tangent construction
on the curve of $\omega$ vs $n_m$ at fixed salt chemical potential, which
imposes equality of the pressure and of the chemical potentials of macroions
and salt in coexisting phases.  As an internal consistency check, it is
possible also to calculate the chemical potentials of the individual microion
species, $\mu_{\pm}=(\partial F/\partial N_{\pm})_{V,N_m,N_{\mp}}$.
Chemical equilibrium between the suspension and reservoir requires
\begin{equation}
\beta(\mu_+-\mu_-)=\ln\left(\frac{n_+}{n_-}\right)
-2~\frac{n_+-n_-}{n_{\mu}}=0,
\label{mudiff}
\end{equation}
which follows from Eq.~(\ref{E02}) and symmetry of $\kappa$ with respect to
interchange of $n_+$ and $n_-$.

\section{Results and Discussion}\label{results}

\begin{figure}
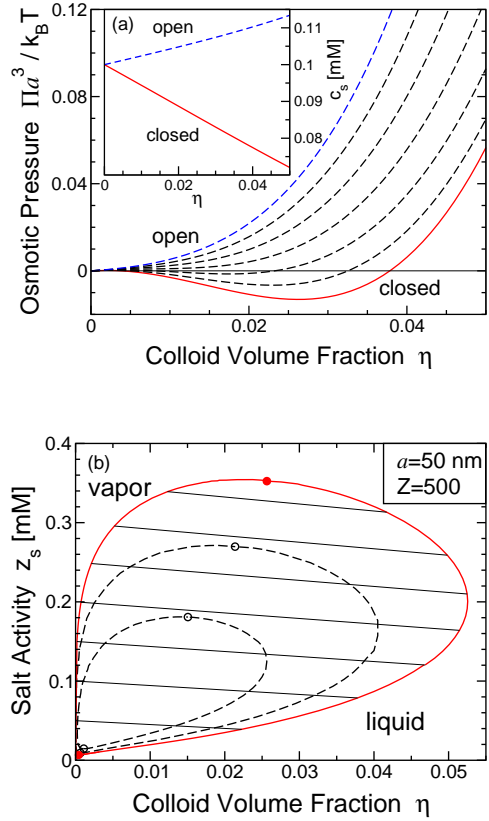

\includegraphics[width=0.8\columnwidth]{donnanp_100_500_new.eps}
\\[6ex]
\includegraphics[width=0.8\columnwidth]{pdlv.100.500.eps}
\caption{\label{press} 
Linear-response theory predictions for osmotic pressure (a) and phase diagram
(b) of aqueous suspensions of macroions of diameter $a=50$ nm and fixed
effective valence $Z=500$ in Donnan equilbrium with a 1:1 microion reservoir
at salt activity $z_s=0.1$ mM.  Perturbation about closed reference system
predicts phase instability [(a), solid curve] and vapor-liquid coexistence
[(b), solid curve].  Perturbation about open reference system (dashed curves)
predicts increased stability with increasing reservoir-to-suspension volume ratio
[$V_r/V=0, 0.2, 0.5, 1, 2, 5, \infty$, bottom to top in (a)] and,
correspondingly, a shrinking binodal [$V_r/V=0$, 0.2, 0.5, largest to smallest
area in (b)].  Instability vanishes and binodal collapses for $V_r/V>1$.
Circles denote critical points and tie lines join coexisting phases.
Inset of (a): salt concentration of suspension $c_s$ with closed (solid) and
open (dashed) reference systems for infinite reservoir ($V_r/V=\infty$).
}
\end{figure}

%Osmotic pressure:
The importance of choosing an appropriate reference system in linearized
theories becomes clear upon examining the osmotic pressure $\Pi=p-2n_rk_BT$
-- the difference between the suspension and reservoir pressures -- at
constant salt activity $z_s=\exp(\beta\mu_s)/\Lambda^3$,
along with the corresponding phase behavior.  
Figure~\ref{press} shows sample predictions of linear-response theory,
with both closed and open reference systems, for macroion radius $a=50$ nm
and fixed effective valence $Z=500$.  While 
the present theory neglects any influence of charge renormalization on $Z$, 
and thus on phase behavior, the results at least qualitatively illustrate the 
significance of the electroneutrality constraint for deionized suspensions.

%Closed reference system:
Linearization about a closed reference system predicts, at sufficiently
low salt activity, a van der Waals loop in the equation of state ($\Pi$ vs 
$\eta$), i.e., a range of volume fractions over which the compressibility is
negative, independent of reservoir size [solid curve of Fig.~\ref{press}(a)].
This unusual spinodal instability, first discovered by van Roij 
\etal~\cite{vRH97,vRDH99,vRE99}, implies separation into macroion-rich 
(liquid) and macroion-poor (vapor) bulk phases~\cite{note4} within the binodal
of the corresponding phase diagram [solid curve of Fig.~\ref{press}(b)]. 
Close examination reveals the instability to be a many-body effect driven by 
the density dependence of both the macroion self-energy and the 
electroneutrality terms in the volume energy 
[last two terms of Eq.~(\ref{E02})].

%Open reference system:
Linearization about an open reference system predicts, in sharp contrast,
a van der Waals loop that narrows, and a liquid-vapor binodal that shrinks, 
with increasing reservoir volume [dashed curves of Fig.~\ref{press}].  
For $V_r>V$, the loop closes entirely and the binodal collapses, the 
compressibility then being strictly positive at all salt concentrations, 
implying a single stable fluid phase.  Such a trend would imply the unphysical
possibility of influencing bulk phase behavior by controlling the exchange 
of microions between suspension and reservoir, i.e., simply adjusting chemical
boundary conditions.  The corresponding salt concentrations in the suspension
$c_s$ (inset to Fig.~\ref{press}(a)) also show qualitatively distinct trends, 
decreasing or increasing with increasing $\eta$ for closed or open reference 
systems, respectively.  The fact that real salt concentrations
are generally lower in the suspension than in the reservoir~\cite{shaw}
indicates a further unphysical property of the open reference system.

\begin{figure}
\includegraphics[width=0.8\columnwidth]{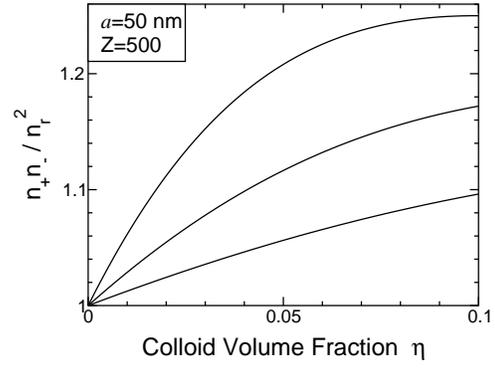}
\vspace*{0.001cm}
\caption{\label{ineq}
Microion density inequality~\cite{deserno02}, $n_+n_-\geq n_r^2$,
tested (and confirmed) at same parameters as Fig.~\ref{press} and
salt activities $z_s=0.1, 0.2, 0.4$ mM (top to bottom).
}
\end{figure}

%Consistency checks:
Accuracy of the linearization approximation can be probed by 
self-consistency checks.
First, the average linearized potential in the suspension can be estimated 
from Eq.~(\ref{Psir}):
\begin{eqnarray}
\beta e|\overline\Psi|&=&4\pi\beta e\frac{N_m}{V'}\int_a^{\infty}{\rm d}r\,
r^2|\Psi(r)|
\nonumber \\[1ex]
&=&\frac{3}{(\kappa a)^2}\frac{Z\lambda_B}{a}\frac{\eta}{1-\eta}.
\label{Psiav}
\end{eqnarray}
For the electrostatic coupling considered here ($Z\lambda_B/a\simeq 7$),
this estimate yields $\beta e|\overline\Psi|<0.3$ along the binodal of
Fig.~\ref{press}(b).  Although the potential is relatively high at the 
surface, $\beta e|\Psi(a)|\simeq 3$, it decays rapidly with distance from
the surface.  Integration of the microion density profiles~\cite{note5} 
for these parameters shows that only 10-20\% of the counterions are in 
the highly nonlinear region, 
$\beta e|\Psi(r)-\bar\Psi|>1$, for salt activities $z_s>0.05~$mM.
By comparison, the artifacts of linearized PB theory demonstrated in 
Refs.~\cite{vongrunberg01,klein01,deserno02,tamashiro03} occur at much 
stronger couplings ($Z\lambda_B/a>18$ and $\beta e|\overline\Psi|>1$),
where nonlinear effects are more significant.
Second, when compared with (salt-free) primitive model 
simulations~\cite{linse00}, linear-response theory~\cite{lu-denton07} 
accurately predicts osmotic pressures for the same strength of electrostatic 
coupling, $Z\lambda_B/a=7.1$ ($Z=40, \lambda_B/a=0.1779$).
Third, the inclusion of first-order nonlinear terms in the effective 
interactions has been shown~\cite{denton06} for these parameters to only 
quantitatively alter the phase diagram, although higher-order nonlinear 
corrections may not necessarily be small.
Fourth, the difference in chemical potentials between microion species
[from Eq.~(\ref{mudiff})] is found to be negligible.
Finally, a rigorous inequality involving the average microion densities in
the suspension~\cite{deserno02}, $n_+n_-\geq n_r^2$, which follows from
Eq.~(\ref{npm1}), is safely satisfied, as demonstrated in Fig.~\ref{ineq}.
Nevertheless, it is conceivable that charge renormalization
could substantially modify the predicted phase behavior, even for such 
a relatively weak coupling.

%Reference system revisited:
Returning to the choice of reference system, the common perturbative origin
of linearized PB and response theories suggests a natural selection criterion.
To minimize the magnitude of the perturbation term in the microion grand
potential [Eq.~(\ref{Omegamu2})], and thereby optimize the accuracy of
linearization, the reference system {\it alone} should reasonably describe
the real suspension.  On this basis, global electroneutrality clearly favors
the closed over the open reference system for a bulk suspension.  
To appreciate why, consider that even monovalent macroions
($Z=1$) would generate a macroscopic charge, which must be neutralized by a 
compensating counterion charge, lest the energy density diverge.  Only a 
reference system whose counterions are confined to the same volume as the 
macroions, with a density hence slaved to the macroions, respects this 
constraint.  
Electroneutrality thus dictates the reference system of a bulk suspension
and decouples phase behavior from boundary conditions.

%Low-dimensional systems:
A caveat to the above conclusions may be required in the case of Donnan
equilibrium of lower-dimensional colloidal assemblies, e.g., monolayers 
(quasi-2D) or clusters (quasi-0D), whose ions are all near the surface 
and in direct contact with a reservoir.  In such systems, electroneutrality 
does not necessarily constrain the majority of counterions to the volume 
occupied by the macroions.  The distribution of counterions between the
macroion region and the reservoir depends instead on the strength of 
macroion-counterion attraction and may determine whether a suspension is 
best modeled by an open or closed reference system.

%Gouy-Chapman length of monolayers:
For a colloidal monolayer (e.g., at an air-water interface or a glass surface),
the range of counterion localization, compared with layer thickness, can be 
roughly quantified by estimating the Gouy-Chapman length $l$, i.e., the 
typical distance to which thermal energy can separate a counterion from the 
monolayer surface by working against electrostatic attraction of the macroions.
Approximating a monolayer as a uniform surface charge density, 
and neglecting electrolyte screening, yields a lower limit of 
$l\simeq(\pi\alpha Z\lambda_B\rho)^{-1}$, where $\alpha$ is the fraction of 
escaped counterions and $\rho$ is the areal density.  
In the dilute limit ($\rho a^2\ll 1$), the Gouy-Chapman length 
can far exceed the monolayer thickness ($l\gg a$), implying weak localization
of counterions by the macroions.
In contrast, a suspension that is macroscopic in all three dimensions, 
{\it always} confines the vast majority of its counterions in bulk,
considering that $l$ is a microscopic length.

%Escape energy of clusters:
For a small cluster of macroions, a natural measure of counterion confinement
is the escape energy, i.e., the energy required to remove a counterion from the
cluster to the surrounding reservoir.  An isolated, spherical cluster of $N_m$
macroions with total charge $\alpha N_mZe$ and volume fraction $\eta$ binds
a counterion near its surface with an escape energy $E_{\rm esc}$ that has an
upper limit $\beta E_{\rm esc}\simeq\alpha(Z\lambda_B/a)N_m^{2/3}\eta^{1/3}$,
again neglecting electrolyte screening.  While counterions are tightly bound
by large, densely charged clusters, they can escape from sufficiently small,
dilutely charged clusters via thermal evaporation into the reservoir.

%Implications:
These considerations suggest that the effective interactions governing
lower-dimensional colloidal assemblies may arise from a 
reference system dependent on particle number and charge density.  A closed 
reference system -- appropriate at high $N_m$, $\eta$, and $Z$ -- clearly
entails density-dependent effective interactions and many-macroion cohesion 
in deionized suspensions.  In contrast, an open reference system -- 
appropriate at lower $N_m$, $\eta$, and $Z$, where the counterions are not 
enslaved by electroneutrality to the macroions -- may yield 
density-independent effective interactions and pair repulsion of macroions.

%Experimental support:
Some support for this view comes from observations of quasi-2D monolayers
in deionized aqueous suspensions~\cite{brunner02,klein02,dobnikar_jpcm03}.
In these experiments, highly charged PS particles were confined to a plane
by light pressure and corralled via optical traps to effective surface charge
densities of ${\cal O}(10^{-5})$ C/m$^2$ or ${\cal O}(10^{-4})~e$/nm$^2$.
Effective pair potentials, obtained by inverting measured radial distribution
functions, exhibited a density-independent Yukawa form at lower densities,
crossing over to density-dependent non-Yukawa behavior at higher densities.
These observations are consistent with salt-dominated linear screening in
dilute monolayers, whose microion population is fixed by the reservoir, and
nonlinear screening above a threshold macroion density.  Colloidal monolayers
confined between glass plates also reportedly can exhibit density-dependent
effective pair interactions~\cite{arauz-lara96}, although interpretation of
such experiments is complicated by uncontrolled microion exchange with the
glass surfaces.

Some evidence for size dependence in the stability of colloidal clusters 
comes from observations of metastable crystallites and macroion gathering
in low-ionic-strength suspensions compressed near glass plates by
external electric fields~\cite{grier97,muramoto-ito97}.
Slow counterion evaporation from the macroions, and the associated crossover
from cohesive many-body to repulsive pair effective interactions, might
qualitatively explain the slow break-up of small macroion aggregates adrift
in deionized water.  For example, the fcc crystallites reported in
Ref.~\cite{grier97} -- characterized by $a=326$ nm, $Z=7300$,
$\lambda_B=0.72$ nm, $c_s<5~\mu$M, and nearest-neighbor separations
$d=1.8~\mu$m -- would correspond to $\beta E_{\rm esc}<0.56\alpha N_m^{2/3}$.
A crystallite of size $N_m={\cal O}(10^2)$ could conceivably break up by
gradually losing a substantial fraction of its counterions to the surrounding
reservoir.  Such a destabilization mechanism also could account for
the observed increase in stability with increasing cluster size, without
invoking any long-range pairwise attractive interaction.
Although a crossover between closed and open reference systems appears
consistent with these experimental observations, a quantitative description 
requires further work.

\section{Conclusions}\label{conclusions}

Summarizing, this work analyzes the influence of the reference system on
the effective electrostatic interactions and phase behavior predicted by
linearized, coarse-grained theories of charged colloids.  For a suspension
in Donnan equilibrium with a microion reservoir, two reference systems are
conceivable: one closed and the other open with respect to microion exchange
with the reservoir.  The constraint of global electroneutrality is shown to
provide an objective physical basis for selecting between these two reference
systems.  From this criterion, and the unification of Poisson-Boltzmann and
response theories, it is concluded that bulk suspensions are properly modeled
within PB theory by expanding about the average potential and average microion
densities within the suspension ({\it not} the reservoir), and within 
response theory by perturbing about a closed reference system, whose counterions 
share the same volume as the macroions.

Dependence of predictions on the reference system need not necessarily
constitute a failing of linearized theories; it merely underscores the
importance of correctly choosing the reference system.  The optimal reference
system identified in Ref.~\cite{deserno02} for PB theory is the same one that
respects electroneutrality in linear-response theory.  The present conclusions
thus broaden and clarify those of previous studies based on PB cell
models~\cite{vongrunberg01,klein01,deserno02,tamashiro03}.
When linearized about a physically consistent reference system, both PB
and response theories predict a spinodal phase instability of highly 
deionized suspensions, assuming a state-independent effective macroion 
charge.  Whether the predicted instability has any relevance to experimentally 
observed anomalies, however, can be decided only by resolving the important 
issues of nonlinear screening and charge renormalization.

Lower-dimensional systems -- from monolayers to clusters -- may be best
described, depending on the strength of macroion-counterion correlations,
by a closed reference system at high charge densities, but an open reference
system at lower charge densities.  The implied transition in effective
interactions and thermodynamic properties may parallel an observed crossover,
as a function of density, in the structure of colloidal monolayers and might
qualitatively explain reports of metastable crystallites.  Further comparisons
with experiment are required to more fully evaluate these speculations.
Future work should seek a more unified, cross-dimensional description of
charged colloids.  Generalization of the coarse-grained approach to richer
models with multiple microion species -- some exchanging with a reservoir
and some trapped -- may have applications to colloids or nanoparticles
confined in pores and to proteins inside cells, where ion channels regulate
exchange of ions across cell membranes.

%------------------------------------------------------------

\begin{acknowledgments}
I thank Alexander Wagner, Sylvio May, and Anne Denton for inspiring discussions.
Parts of this work were supported by National Science Foundation Grant
No.~DMR-0204020.
\end{acknowledgments}

%\end{multicols}

%\appendix*

%%%%%%%%%%%%%%%%%%%%%%%%%%%%%%%%%%%%%%%%%%%%%%%%%%%%%%%%%%%%%%%%%%%%%%%%%%%
%%%%%%%                       REFERENCES
%%%%%%%%%%%%%%%%%%%%%%%%%%%%%%%%%%%%%%%%%%%%%%%%%%%%%%%%%%%%%%%%%%%%%%%%%%%

%%%%%%%%%%%%%%%%%%%%%%%%%%%%%%%%%%%%%%%%%%%%%%%%%%%%%%%%%%%%%%%%%%%%%%%%%%%
%%%%%%%%                       FIGURES
%%%%%%%%%%%%%%%%%%%%%%%%%%%%%%%%%%%%%%%%%%%%%%%%%%%%%%%%%%%%%%%%%%%%%%%%%%%

%  use \protect\command{}  if you have to use a command \command{}
%  which has an argument

% in order to put the figures into the text you have to activate
% the line with ``\input{psfig}'' as well

%\unitlength1mm


\begin{references}

% chose byline symbols in the order *, \dag, \ddag, \S, **, \dag\dag
%\bibitem[*]{byline1}

\bibitem{evans99}
D.~F.~Evans and H.~Wennerstr\"om, {\it The Colloidal Domain}, 2$^{\rm nd}$ ed.
(Wiley-VCH, New York, 1999).

\bibitem{tata92}
B.~V.~R.~Tata, M.~Rajalakshmi, and A.~K.~Arora, \PRL {\bf 69}, 3778 (1992).

\bibitem{ise94}
K.~Ito, H.~Yoshida, and N.~Ise, Science {\bf 263}, 66 (1994).

\bibitem{tata97}
B.~V.~R.~Tata, E.~Yamahara, P.~V.~Rajamani, and N.~Ise,
\PRL {\bf 78}, 2660 (1997).

\bibitem{ise99}
N.~Ise, T.~Konishi, and B.~V.~R.~Tata, Langmuir {\bf 15}, 4176 (1999).

\bibitem{harada99}
T.~Harada, H.~Matsuoka, T.~Ikeda, and H.~Yamaoka,
Langmuir {\bf 15}, 573 (1999).

\bibitem{grier97}
A.~E.~Larsen and D.~G.~Grier, Nature {\bf 385}, 230 (1997).

\bibitem{muramoto-ito97}
T.~Muramoto, K.~Ito, and H.~Kitano, J. Am. Chem. Soc.
{\bf 119}, 3592 (1997).

\bibitem{DLVO}
B.~V.~Derjaguin and L.~Landau, Acta Physicochimica (URSS) {\bf 14},
633 (1941);
E.~J.~W.~Verwey and J.~T.~G.~Overbeek, {\it Theory of the Stability of
Lyophobic Colloids} (Elsevier, Amsterdam, 1948).

\bibitem{neu99}
J.~C.~Neu, \PRL {\bf 82}, 1072 (1999).

\bibitem{palberg94}
T.~Palberg and M.~W\"urth, \PRL {\bf 72}, 786 (1994);
B.~V.~R.~Tata and A.~K.~Arora, {\it ibid} {\bf 72}, 787 (1994).

\bibitem{belloni00}
L.~Belloni, \JPCM {\bf 12}, R549 (2000).

\bibitem{schmitz99}
K.~S.~Schmitz, \PCCP {\bf 1}, 2109 (1999).

\bibitem{allahyarov-damico98}
E.~Allahyarov, I.~D'Amico, and H.~L\"owen, \PRL {\bf 81}, 1334 (1998).

\bibitem{linse-lobaskin99}
P.~Linse and V.~Lobaskin, \PRL {\bf 83}, 4208 (1999).

\bibitem{lobaskin-linse99}
V.~Lobaskin and P.~Linse, \JCP {\bf 111}, 4300 (1999).

\bibitem{linse99}
P.~Linse, \JCP {\bf 110}, 3493 (1999).

\bibitem{linse00}
P.~Linse, \JCP {\bf 113}, 4359 (2000).

\bibitem{rescic-linse01}
J.~Re\v{s}\v{c}i\v{c} and P.~Linse, \JCP {\bf 114}, 10131 (2001).

\bibitem{lobaskin-linse01}
V.~Lobaskin, A.~Lyubartsev, and P.~Linse, \PR E {\bf 63}, 020401(R) (2001).

\bibitem{lobaskin-qamhieh03}
V.~Lobaskin and K.~Qamhieh, J. Phys. Chem. B {\bf 107}, 8022 (2003).

\bibitem{hynninen05}
A.-P.~Hynninen, M.~Dijkstra, and A.~Z.~Panagiotopoulos,
\JCP {\bf 123}, 84903 (2005).

\bibitem{hynninen07}
A.-P.~Hynninen and A.~Z.~Panagiotopoulos, \PRL {\bf 98}, 198301 (2007).

\bibitem{vRH97}
R.~van Roij and J.-P.~Hansen, \PRL {\bf 79}, 3082 (1997).

\bibitem{vRDH99}
R.~van Roij, M.~Dijkstra, and J.-P.~Hansen, \PR E {\bf 59}, 2010 (1999).

\bibitem{vRE99}
R.~van Roij and R.~Evans, \JPCM {\bf 11}, 10047 (1999).

\bibitem{zoetekouw_pre06}
B.~Zoetekouw and R.~van Roij, \PR E {\bf 73}, 21403 (2006).

\bibitem{warren00}
P.~B.~Warren, \JCP {\bf 112}, 4683 (2000); \JPCM {\bf 15}, S3467 (2003);
\PR E {\bf 73}, 011411 (2006).

\bibitem{chan85}
B.~Beresford-Smith, D.~Y.~C.~Chan, and D.~J.~Mitchell, \JCIS {\bf 105},
216 (1985).

\bibitem{chan01}
D.~Y.~C.~Chan, P.~Linse, and S.~N.~Petris, Langmuir {\bf 17}, 4202 (2001).

\bibitem{silbert91}
M.~J.~Grimson and M.~Silbert, \MP {\bf 74}, 397 (1991).

\bibitem{denton99}
A.~R.~Denton, \JPCM {\bf 11}, 10061 (1999).

\bibitem{denton00}
A.~R.~Denton, \PR E {\bf 62}, 3855 (2000).

\bibitem{denton04}
A.~R.~Denton, \PR E {\bf 70}, 31404 (2004).

\bibitem{denton06}
A.~R.~Denton, \PR E {\bf 73}, 41407 (2006).

\bibitem{int-eq}
For a recent integral-equation theory approach, see
J.~A.~Anta and S.~Lago, \JCP {\bf 116}, 10514 (2002);
V.~Morales, J.~A.~Anta, and S.~Lago, Langmuir {\bf 19}, 475 (2003).

\bibitem{donnan11}
F.~G.~Donnan, Z. Elektrochem. Angew. Phys. Chem. {\bf 17}, 572 (1911);
English translation: J. Membrane Sci. {\bf 100}, 45 (1995).

\bibitem{donnan24}
F.~G.~Donnan, Chem. Rev. {\bf 1}, 73 (1924).

\bibitem{vongrunberg01}
H.~H.~von Gr\"unberg, R.~van Roij, and G.~Klein, \EPL {\bf 55}, 580 (2001).

\bibitem{klein01}
R.~Klein and H.~H.~von Gr\"unberg, Pure Appl. Chem. {\bf 73}, 1705 (2001).

\bibitem{deserno02}
M.~Deserno and H.~H.~von Gr\"unberg, \PR E {\bf 66}, 011401 (2002).

\bibitem{tamashiro03}
M.~N.~Tamashiro and H.~Schiessel, \JCP {\bf 119}, 1855 (2003).

\bibitem{trizac03}
G.~T\'ellez and E.~Trizac, \JCP {\bf 118}, 3362 (2003).

\bibitem{alexander84}
S.~Alexander, P.~M.~Chaikin, P.~Grant, G.~J.~Morales, and P. Pincus,
\JCP {\bf 80}, 5776 (1984).

\bibitem{levin98}
Y.~Levin, M.~C.~Barbosa, and M.~N.~Tamashiro, \EPL {\bf 41}, 123 (1998).

\bibitem{tamashiro98}
M.~N.~Tamashiro, Y.~Levin, and M.~C.~Barbosa, \EPJ B {\bf 1}, 337 (1998).

\bibitem{levin01}
A.~Diehl, M.~C.~Barbosa, and Y.~Levin, \EPL {\bf 53}, 86 (2001).

\bibitem{levin03}
Y.~Levin, E.~Trizac, and L.~Bocquet, \JPCM {\bf 15}, S3523 (2003).

\bibitem{levin04}
A.~Diehl and Y.~Levin, \JCP {\bf 121}, 12100 (2004).

\bibitem{zoetekouw_prl06}
B.~Zoetekouw and R.~van Roij, \PRL {\bf 97}, 258302 (2006).

\bibitem{dobnikar_epl03}
J.~Dobnikar, R.~Rzehak, and H.~H.~von Gr\"unberg, \EPL {\bf 61}, 695 (2003).

\bibitem{rowlinson84}
J.~S.~Rowlinson, \MP {\bf 52}, 567 (1984).

\bibitem{hansen-lowen00}
J.-P.~Hansen and H.~L\"owen, Annu. Rev. Phys. Chem. {\bf 51}, 209 (2000).

\bibitem{likos01}
C.~N.~Likos, Phys. Rep. {\bf 348}, 267 (2001).

\bibitem{levin02}
Y.~Levin, Rep. Prog. Phys. {\bf 65}, 1577 (2002).

\bibitem{zvelindovsky07}
A.~R.~Denton, in {\it Nanostructured Soft Matter: Experiment, Theory, Simulation
and Perspectives}, ed. A.~V.~Zvelindovsky (Springer, Dordrecht, 2007).

\bibitem{lowen98}
H.~L\"owen and E.~Allahyarov, \JPCM {\bf 10}, 4147 (1998).

\bibitem{goulding99}
D.~Goulding and J.-P.~Hansen, \EPL {\bf 46}, 407 (1999).

\bibitem{note1}
In this paper, the terms ``electrochemical potential" and
``chemical potential" are synonymous.

\bibitem{HM}
J.-P.~Hansen and I.~R.~McDonald, {\it Theory of Simple Liquids},
2nd ed.~(Academic, London, 1986).

\bibitem{lowen93}
H.~L\"owen, J.-P.~Hansen, and P.~A.~Madden, \JCP {\bf 98}, 3275 (1993).

\bibitem{note2}
The basic equivalence between response theory and PB theory extends to
nonlinear response.  See Refs.~\cite{denton04} and \cite{lowen98}.

\bibitem{note3}
All partial derivatives here are taken {\it without} fixing the reference
variables (potential and microion densities), corresponding to the second
pressure definition in Ref.~\cite{deserno02} [Eq.~(22) therein].

\bibitem{lu-denton07}
B.~Lu and A.~R.~Denton, \PR E {\bf 75}, 061403 (2007).

\bibitem{shaw}
D.~J.~Shaw, {\it Introduction to Colloid and Surface Chemistry},
4$^{\rm th}$ ed. (Butterworth-Heinemann, Oxford, 1992), pp. 42-43.

\bibitem{note5}
Integrating the linearized microion density profiles (Eq.~(47) of 
Ref.~\cite{denton00}) yields the fraction of microions within a 
radius $r$ of the macroion centers for microion concentrations
$x_{\pm}=N_{\pm}/N_{\mu}$: 
\begin{eqnarray}
\hspace*{0.7cm}f_{\pm}(r)&=&x_{\mp}\frac{2\eta}{1-\eta}
\left(\frac{r^3}{a^3}-1\right)\pm(x_+-x_-)
\nonumber \\
&\times&\left(1-\frac{1+\kappa r}{1+\kappa a}e^{-\kappa(r-a)}\right),
\quad r/a<1/\eta^{1/3}.
\nonumber
\end{eqnarray}

\bibitem{note4}
For the selected parameters, the vapor-liquid binodal is well separated
from the liquid-solid phase boundary (see Ref.~\cite{lu-denton07}).

\bibitem{brunner02}
M.~Brunner, C.~Bechinger, W.~Strepp, V.~Lobaskin, and H.~H.~von Gr\"unberg,
\EPL {\bf 58}, 926 (2002).

\bibitem{klein02}
R.~Klein, H.~H.~von Gr\"unberg, C.~Bechinger, M.~Brunner, and V.~Lobaskin,
\JPCM {\bf 14}, 7631 (2002).

\bibitem{dobnikar_jpcm03}
J.~Dobnikar, Y.~Chen, R.~Rzehak, and H.~H.~von Gr\"unberg,
\JPCM {\bf 15}, S263 (2003).

\bibitem{arauz-lara96}
M.~D.~Carbajal-Tinoco, F.~Castro-Rom\'an, and J.~L.~Arauz-Lara,
\PR E {\bf 53}, 3745 (1996).

\end{references}
\end{document}